%% file: bare_conf.tex
\documentclass[conference]{IEEEtran}
\input{misc_packages}
\usepackage[noadjust]{cite}

\begin{document}

\title{High Data Rate Near-Ultrasonic Communication with Consumer Devices}

\author{\IEEEauthorblockN{Gizem Tabak$^1$, Xintian Eddie Lin$^2$, and Andrew C. Singer$^1$}
\IEEEauthorblockA{$^1$Department of Electrical and Computer Engineering\\
	University of Illinois at Urbana-Champaign, Urbana, IL, USA\\
	$^2$Intel Corporation, Santa Clara, CA, USA\vspace{-0.2cm}}
	}

\maketitle

\begin{abstract}
Automating device pairing and credential exchange in consumer devices reduce the time users spend with mundane tasks and improve the user experience. Acoustic communication is gaining traction as a practical alternative to Bluetooth or Wi-Fi because it can enable quick and localized information transfer between consumer devices with built-in hardware. However, achieving high data rates ($>$1 kbps) in such systems has been a challenge because the systems and methods chosen for communication were not tailored to the application. In this work, a high data rate, near-ultrasonic communication (NUSC) system is proposed to transfer personal identification numbers (PINs) to establish a connection between consumer laptops using built-in microphones and speakers. The similarities between indoor near-ultrasonic and underwater acoustic communication (UWAC) channels are identified, and appropriate UWAC techniques are tailored to the NUSC system. The proposed system uses the near-ultrasonic band at 18-20 kHz, and employs coherent modulation and phase-coherent adaptive equalization. The capability of the proposed system is explored in simulated and field experiments that span different device orientations and distances. The experiments demonstrate data rates of 4 kbps over distances of up to 5 meters, which is an order of magnitude higher than the data rates reported with similar systems in the literature. 
\end{abstract}

\IEEEpeerreviewmaketitle

\section{Introduction}
With ambient intelligence and Internet-of-Things (IoT) on the rise, occupying the user with mundane tasks such as credential entry and device pairing becomes less desirable. 
Conventionally, most of the information exchange between consumer devices happens through electromagnetic radio-frequency (RF) links such as Wi-Fi or Bluetooth. These technologies provide valuable connectivity and interaction capabilities, but they are subject to standardized rules and protocols. Moreover, they usually require user input and control for initial pairing and activation. In applications such as network credential exchange, user authentication or proximity-based content sharing, a few hundred bits of data are needed to be transmitted quickly and locally. Airborne acoustic communication (AAC) is steadily gaining significance in those scenarios \cite{gupta2020development}, \cite{lee2020reliable}.


The human ear can hear acoustic frequencies between 20 Hz and 20 kHz. However, the ability to hear higher frequencies diminishes with age. Any frequency beyond 18 kHz is rarely audible for anyone over 18 \cite{stelmachowicz1989normative}. Many consumer devices, on the other hand, have audio systems that have sampling rates of 44.1 kHz or above, and they are capable of operating in the near-ultrasonic range of 18-22 kHz \cite{Hanspach2013}. The studies on the effects of airborne ultrasound on human hearing does not suggest an inaudible application using standard computer hardware potentially being unsafe \cite{fletcher2018effects, fletcher2018public} (See Appendix). 
Therefore, a high data rate, near-ultrasonic communication (NUSC) system implemented with the built-in microphones and speakers of consumer devices can provide a safe and inexpensive method for establishing connections quickly between devices.

Three main challenges arise from using the built-in microphones and speakers of consumer devices for high data rate NUSC: First, the frequency response of the built-in hardware in the near-ultrasonic range is usually nonflat \cite{Hanspach2013, Lee2015, lee2020reliable}. Second, the indoor acoustic channel is frequency selective with severe intersymbol interference (ISI) at high rates due to the hardware characteristics and multipath. Finally, the channel can be time-varying because of the movements of users and devices.

Previous studies borrowed methods from radio communication and proposed using noncoherent modulation with symbol durations longer than channel delay spread to mitigate the deleterious effects of NUSC channel. However, the propagation speed of acoustic waves in air is much slower than radio waves, causing a longer delay spread. Therefore, symbols longer than channel delay spread yields data rates on the order of tens of bits per second (bps) in NUSC systems. 

Nonflat frequency response, severe ISI, and time variations are also present in a typical underwater acoustic communication (UWAC) channel \cite{stojanovic2009underwater}. It is demonstrated in UWAC and other related acoustic communications literature that understanding channel properties and tailoring communication methods to address particular challenges may provide significant improvements in data rates \cite{stojanovic2009underwater, riedl2014towards, singer2016mbps, tabak2019ultrasonic, tabak2020video}. 


In this work, 18-20 kHz band of the built-in speakers and microphones of consumer laptops are used for high rate NUSC. The properties and challenges of NUSC channel are presented by highlighting the similarities between UWAC and NUSC channels. An adaptive, phase-coherent decision feedback equalizer (PC-DFE) \cite{stojanovic1994phase} is tailored and used in the proposed system to overcome the three aforementioned challenges. The novel use of PC-DFE at the receiver of the NUSC system enables high data rate communication with phase-coherent modulation. To examine the capabilities of the system, simulated and physical experiments are conducted in various conference room settings, with transmission distances up to 5 meters. The experiments demonstrated data rates of 4 kbps, which is an order of magnitude higher than the data rates reported with consumer laptops and mobile devices in the literature.

In Section \ref{sec:related-work}, the related works in the AAC literature is given. In Section \ref{sec:challenges}, the three main challenges of NUSC is explained, and the similarities between NUSC and UWAC channels are highlighted. In Section \ref{sec:system}, the system proposed to overcome these challenges is introduced. In Section \ref{sec:experiments}, the simulated and physical experiments, and their results are presented. Finally, the paper is concluded in Section \ref{sec:conclusion}. 

\section{Related Work}
\label{sec:related-work}
AAC has been used for sending information between devices at least since 1996 \cite{gerasimov1996thesis}. In \cite{lopes2001aerial, Matsuoka2008, Yun2010}, audible sound is used to embed communication signals and transfer information. In \cite{Holm2005}, Holm \textit{et al.} envisioned 100 bps data rates with range varying from 12 to 60 meters for a hypothetical ultrasonic communication system that would operate at 40 kHz. Since then, several studies proposed building specialized ultrasonic systems that can operate at higher frequencies in the hundreds of kHz range and could achieve rates up to a few hundred kbps over distances varying from a few centimeters to a few meters \cite{Li2009, Jiang2017, jiang2018indoor}.     

More recently, there has been a focus on using the near-ultrasonic band that is already available in many consumer devices such as laptops, TVs and mobile phones. A comparison of the data rates achieved in these works with package error rates varying from 0.008 to 0.1 can be seen in Table \ref{tab:nearUS-lit}. References \cite{lee2020reliable, Lee2015, Ka2016} employed chirp-based modulation, and \cite{Hanspach2013, Getreuer2018} used coherent modulation. None of these works included an equalizer at the receiver. The methods that were chosen for communication resulted in low data rates.

\begin{table}[h]
\caption{Several studies used near-ultrasonic band of consumer devices for sending information over various distances}
    \label{tab:nearUS-lit}
    \centering
    \begin{tabular}{|c|l|l|l|}
    \hline
                              & \textbf{Device}       & \textbf{Data Rate} & \textbf{Distance} \\\hline
        \cite{lee2020reliable} & PC to smartphone & 21 bps\footnotemark & 5-8 m\\
        \cite{Hanspach2013}   & PC to PC        & 20 bps   & 19.7 m  \\
        \cite{Lee2015}        & Speaker to smart device & 16 bps   & 25 m  \\
        \cite{Ka2016}         & TV to smartphone & 15 bps & 2.7 m  \\
        \cite{Getreuer2018}   & Smartphone to smartphone & 94.5 bps & 2 m \\\hline
    \end{tabular}
\end{table}
\footnotetext{Data rates and BERs were not quantified in the study, but the typical quaternary symbol duration was given as 96 ms, suggesting highest possible rate of 21 bps.}
\section{Channel Properties}
\label{sec:challenges}

\subsection{Nonflat Frequency Response}
In a digital communications system, the information is sent from transmitter to receiver in the form of concatenated (and shaped) symbols. The frequency content of the received signal is affected by the frequency response of the speakers of the transmitter device and microphones of the receiver device. For consumer devices, the frequency response of the built-in hardware in the near-ultrasonic range is usually nonflat \cite{Hanspach2013, lee2020reliable}. The nonflat frequency response of the built-in hardware causes dispersion, and the symbol waveforms interfere with each other. This may impede successful recovery of the symbols at the receiver.

In UWAC, the path loss is frequency-dependent because the acoustic absorption in water changes significantly with signal frequency \cite{stojanovic2009underwater}. This results in an inherently nonflat channel response.

\subsection{Multipath Propagation}
In a reverberant indoor space, reflections from the walls and objects cause multipath propagation. The delay spread of an indoor acoustic channel measured in an office space with devices placed 5 m apart could span 70 ms \cite{Lee2015}. Designing symbols longer than channel delay spread yields data rates on the order of tens of bps. If symbols shorter than the channel delay spread are used without any compensation, each symbol arrives at the receiver multiple times at different instances, interfering with and possibly destroying each other.

In UWAC, multipath propagation stems from sound ducting through the non-constant sound-speed profile of the water column, as well as reflections at the surface and bottom, and other scatterers. The channel delay spread can span many symbols with shorter durations at high data rates \cite{stojanovic2009underwater}. Although the refraction of sound in air is not a major contributor to multipath propagation in a typical room, reflections result in long delay spread in both channels and cause severe ISI at high data rates.
\subsection{Time Variations}
In indoor acoustic communication, the transmitted signal is reflected from the walls, objects and users in the environment. The movement of the reflectors and also the movements of the transmitter and receiver devices can cause frequency and phase shifts and time variations in the channel. These effects may impede successful transmission, especially when phase-coherent modulation is employed. 

Similar time variations are also observed in UWAC channel because of reflector, transmitter and receiver platform motion as well as inherent variations due to internal and surface waves.

\section{Communication System}
\label{sec:system}
With the nonflat frequency response, ISI and time variability, the nonidealities of the indoor NUSC are similar to the nonidealities of the UWAC channel. Coherent modulation together with adaptive equalization have successfully been used in various acoustic communication applications to overcome these challenges and achieve high data rates \cite{stojanovic1994phase, riedl2014towards, singer2016mbps, tabak2019ultrasonic, tabak2020video}. Therefore, it is promising to use coherent modulation and adaptive equalization to compensate for the effects of these nonidealities and to increase data rates in NUSC.
\subsection{Signal Design}
In NUSC, the available frequency band is limited from below by the maximum audible frequency and from above by the frequency response of the audio hardware. The data rate of the system is then given by
\begin{equation}
    R = f_b \log_2 M
\end{equation}
where $f_b$ is the symbol rate and $M$ represents the order of modulation and indicates the number of possible values the transmitted symbols may take. This is in similar form to the Shannon channel capacity given by $W\log(1+SNR)$, for bandwidth $W$ and signal to noise ratio $SNR$ and which provides an upper bound on the performance of any such communication system. The maximum available symbol rate $f_b$ is limited by the bandwidth of the communication channel. Hence, in order to achieve high data rates, the available bandwidth should be used as efficiently as possible. In this work, quadrature amplitude modulation (QAM) is used due to its potential for high spectral efficiency.  
The bitstream obtained from the source 
are grouped in $\log_2 M$ bits and mapped into $N$ transmit symbols, denoted by $\{x_1,\dots,x_N\}$ where $x_i\in \mathbb{C}$. 
The achievable constellation order $M$ is limited by operational SNR.
The symbols are upsampled by $L=\frac{f_s}{f_{b}}$, where $f_s\geq 2f_b (1+\beta)$ is the sampling frequency of the audio system, and shaped with a root-raised cosine filter $p(t)$ with roll-off factor $\beta$, resulting in the complex baseband waveform
\begin{equation}
    \tilde{x}(t)=\sum_{k=1}^{N}x_kp(t-kT_b)
\end{equation}
where $T_b=\tfrac{1}{f_b}$ is the symbol period. Then, the baseband waveform is mixed with a sinusoidal carrier at the center frequency of the transmission band, $f_c$. The passband signal
\begin{equation}\label{eq:1-2-passband-sig}
    x(t) = \mathcal{R}e\left\{\sum_{k=1}^{N}x_kp(t-kT_b)e^{j2\pi f_c t}\right\}
\end{equation}
is preceded by a pilot waveform and sent through the channel. A $16^{th}$ order Frank code is used as the pilot waveform to establish synchronization at the receiver. Frank code is preferred due to its synchronization performance in environments that cause multipath propagation \cite{frank1963polyphase}. 
\subsection{Equalizer}
For channels with ISI, the optimum receiver (optimum in the sense that it minimizes sequence error probability) is matched filter followed by a maximum likelihood sequence estimator (MLSE). However, the complexity of MLSE grows exponentially with the channel impulse response length. This makes it an almost intractable solution for NUSC channels. 
Therefore, a suboptimal alternative to MLSE with lower complexity is more appropriate for NUSC channels. One such alternative is the decision feedback equalizer (DFE). The DFE consists of two parts: A feedforward filter and a feedback filter; the feedforward filter operates on the received signal, and feedback filter ``feeds back'' the past symbol decisions. 
One way to view the operation of the DFE is to consider the feedforward filter as providing an MMSE estimate of the symbol assuming that all post-cursor ISI can be ignored (canceled out) and the feedback filter is then designed to cancel this causal ISI assuming perfect feedback. The channel in NUSC is typically unknown and possibly time varying. Therefore, the filters are updated with a training sequence using an adaptive gradient descent or least squares algorithm.

The pre-cursor portion of the feedforward filter determines the delay  introduced into the system. For a quick information exchange application, low delay is desired. Therefore, it is important to consider tolerable delay when determining the span of the feedforward filter.

Since QAM is phase-coherent, phase information is also needed in the recovery of the symbol. Recovering both symbol phase information and symbol timing correctly at the receiver can be a challenge for high data rate applications with possibly time varying channels. 
In order to cope with mild variations in the NUSC channel, a phase-tracking component (e.g. a phase-locked loop) is inserted into the DFE \cite{stojanovic1994phase} so that the DFE will not only update the filters but also correct for the estimated phase and timing offsets of the received signal. 

\section{Experiments}
\label{sec:experiments}
Three sets of experiments were performed to demonstrate the capabilities of the proposed system. The first set of experiments were simulated using room impulse responses. The second and third sets of experiments were performed in various conference rooms with people (experimenters), objects (tables, chairs, TVs, whiteboards, office supplies) and devices that operate at the ultrasonic frequencies (occupancy sensors).

\begin{figure}[t]
     \centering\includegraphics[width=.85\columnwidth]{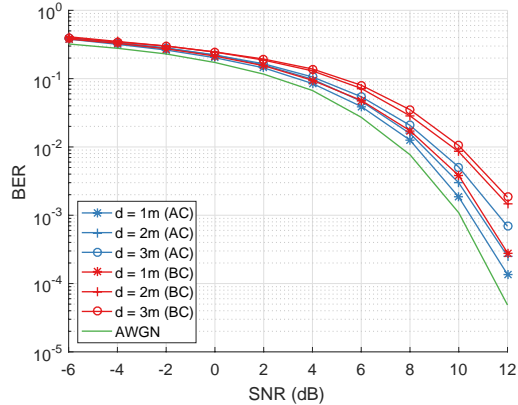}
     \caption{Simulations performed with room impulse responses measured at different microphone-speaker distances}
     \label{fig:RIR_sim_BERvsSNR}
\vspace{-0.2cm}
\end{figure}

In all the experiments, the passband signal was centered at $f_c=$19 kHz and the symbol rate was $f_b=$ 2 kHz. Quadrature phase shift keying (QPSK) was used for modulation, yielding data rates of 4 kbps. A root-raised cosine filter with $0.3$ roll-off factor was used for symbol shaping. Each transmission packet consisted of the pilot signal, a 0.3 second long guard interval and the symbol waveform that contained 5120 data bits and 32 cyclic redundancy check (CRC-32) bits. At the receiver, 30\% of the symbols were used for learning the DFE filter coefficients (data-directed mode). In the remaining, the symbol estimates were used to update the filter coefficients (decision-directed mode). The feedforward filter had 8 noncausal and 20 causal taps, and the feedback filter had 80 taps. The number of noncausal feedforward filter taps was kept low due to the delay constraints. 

\subsection{Simulations}\label{ssec:2-task1-sims}
In order to evaluate the performance limits of the proposed method and to compare the results of different configurations with different transmitter and receiver positions in a room, a series of simulated experiments were performed with the room impulse responses obtained from the MARDY dataset \cite{wen2006evaluation}. The transmission signal was generated according to \eqref{eq:1-2-passband-sig}, convolved with the room impulse response, and different levels of additive white Gaussian noise (AWGN) were added to obtain the simulated received signal with different SNR levels.

The simulated bit error rate (BER) and SNR curves for two configurations are displayed in Fig. \ref{fig:RIR_sim_BERvsSNR}. AC represents the configuration where the microphone and the speaker were placed in the center of the room, and BC represents one where the microphone and the speaker were in the opposite sides of the room near the corners. The solid green line in Fig. \ref{fig:RIR_sim_BERvsSNR} displays the communication performance with ideal AWGN channel. The decoding performance with PC-DFE was within 1-3 dB of the AWGN channel. 
Therefore, the simulation results showed promise towards 4 kbps data rates using near-ultrasonic frequencies. 

\begin{figure}[t]
\begin{minipage}[b]{1.0\linewidth}
  \centering
  \centerline{\includegraphics[width=8.5cm]{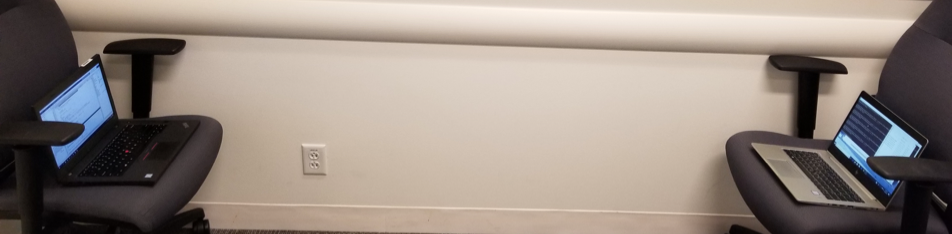}}
  \centerline{(a) Setup 1}\medskip
\end{minipage}
\begin{minipage}[b]{.48\linewidth}
  \centering
  \centerline{\includegraphics[width=3.8cm]{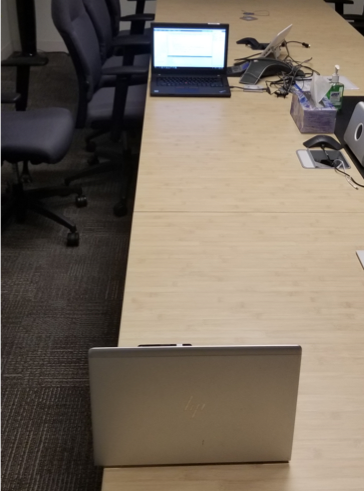}}
  \centerline{(b) Setup 2}\medskip
\end{minipage}
\hfill
\begin{minipage}[b]{0.48\linewidth}
  \centering
  \centerline{\includegraphics[width=4.0cm]{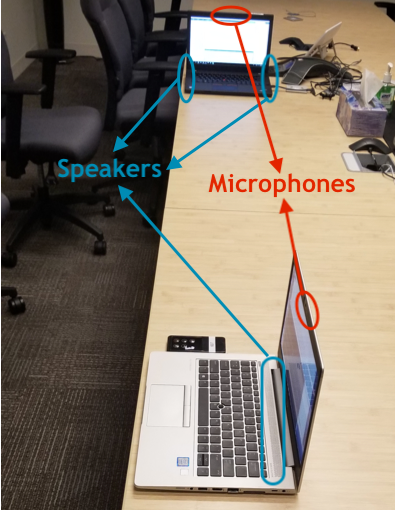}}
  \centerline{(c) Setup 3}\medskip
\end{minipage}
\caption{Two laptops placed at 2 meters distance with different orientations.}
\label{fig:res}
\vspace{-0.2cm}
\end{figure}
\subsection{Device Orientation}\label{ssec:orient}
The orientation of the communicating devices in a conference room affects the channel characteristics because it determines the existence of a line-of-sight path as well as the strength of direct and reflected paths \cite{Lee2015}. An experimental setup was designed to evaluate the effect of device orientation. Two communicating laptops (Device 1 - HP EliteBook 830 and Device 2 - Lenovo T460p) were placed at 2 meters apart with different orientations and surroundings (Fig. \ref{fig:res}). The TX sound level was varied from 50\% to 100\%, and the experiments were repeated 5 times for each configuration.

\begin{figure}[t]
     \centering
     \begin{subfigure}[t]{0.32\columnwidth}
         \centering\includegraphics[width=\textwidth]{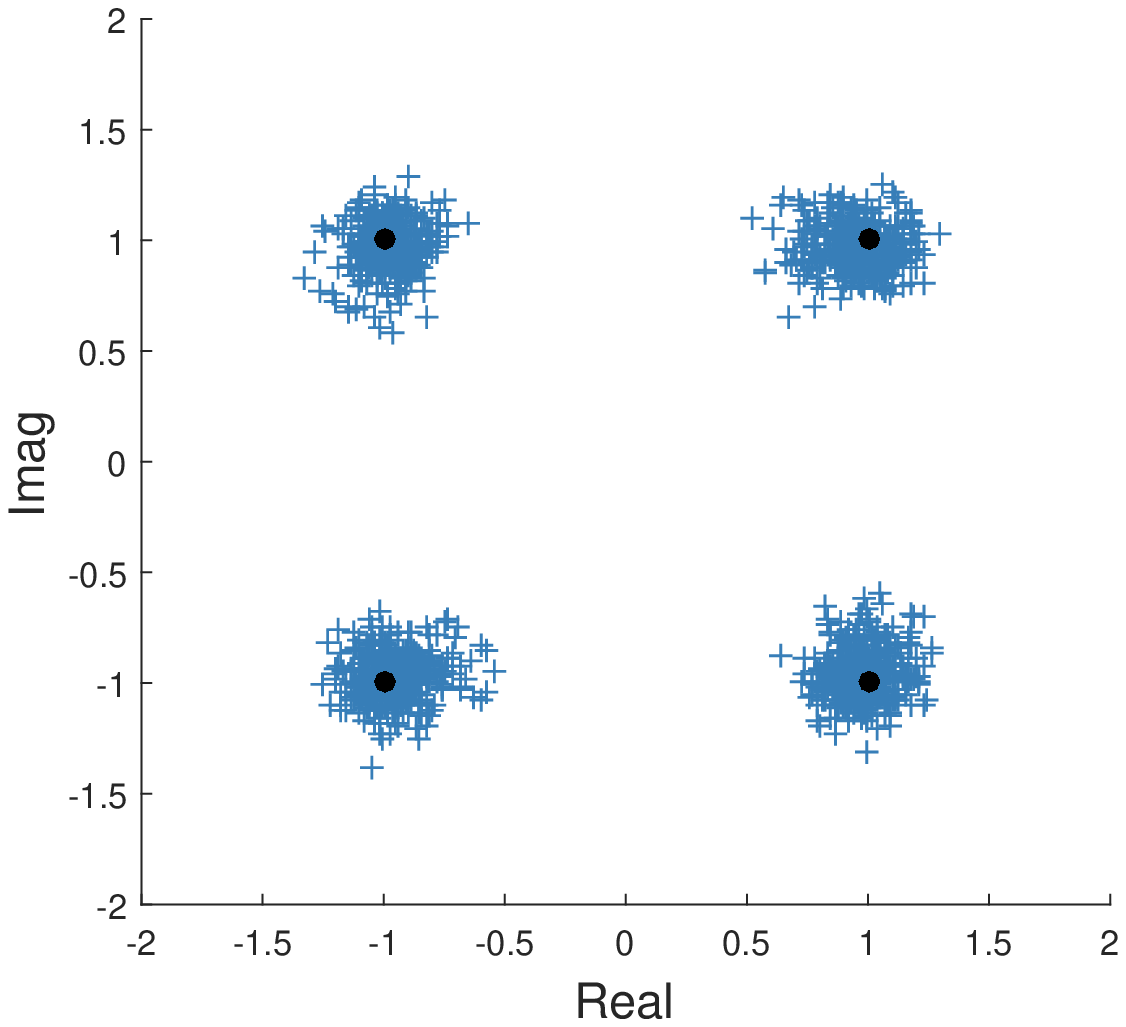}
         \caption{Setup 1}\label{fig:const-sett1}
     \end{subfigure}
     \begin{subfigure}[t]{0.32\columnwidth}
         \centering\includegraphics[width=\textwidth]{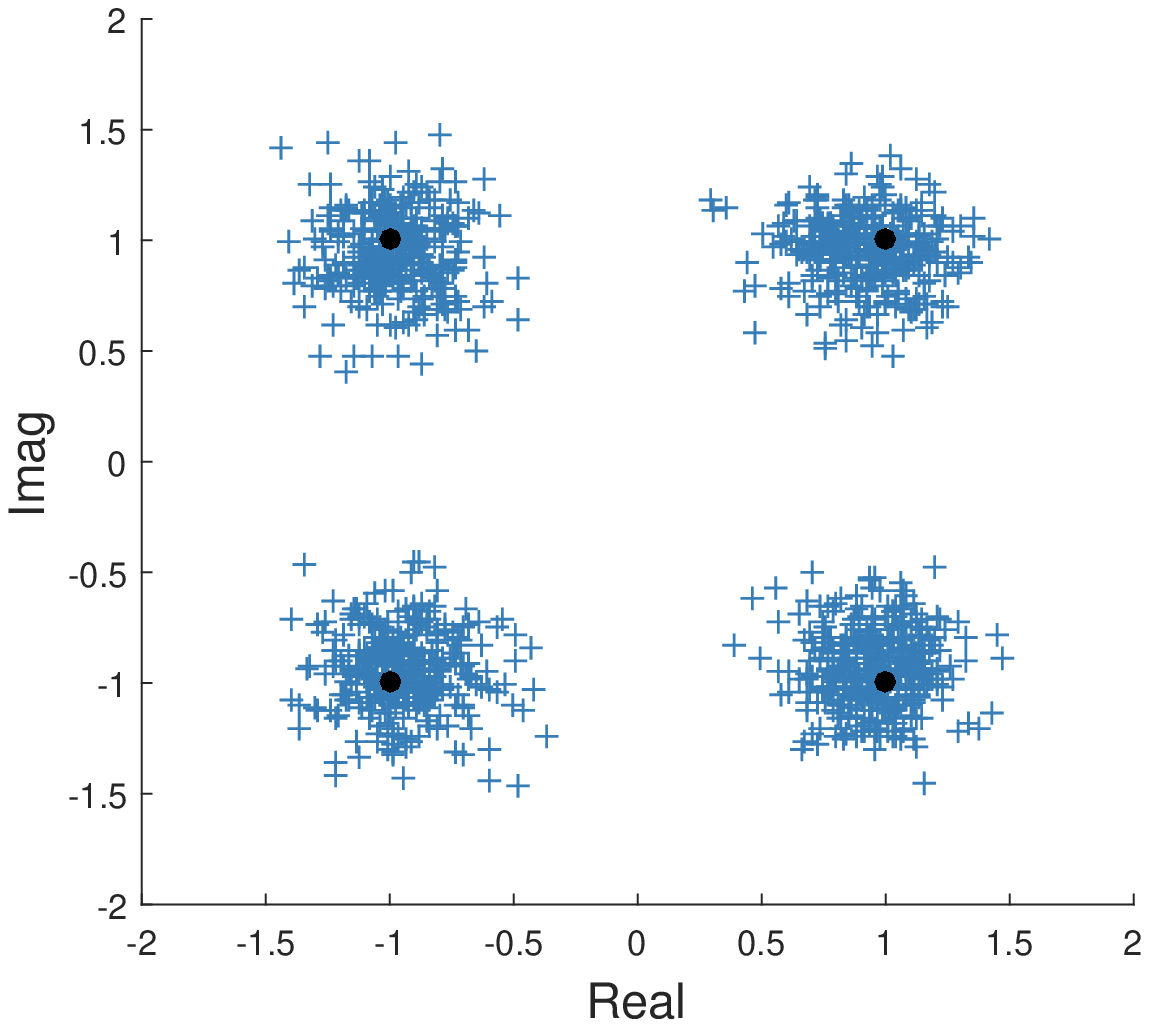}\label{fig:const-set2}
         \caption{Setup 2}
     \end{subfigure} 
     \begin{subfigure}[t]{0.32\columnwidth}
         \centering\includegraphics[width=\textwidth]{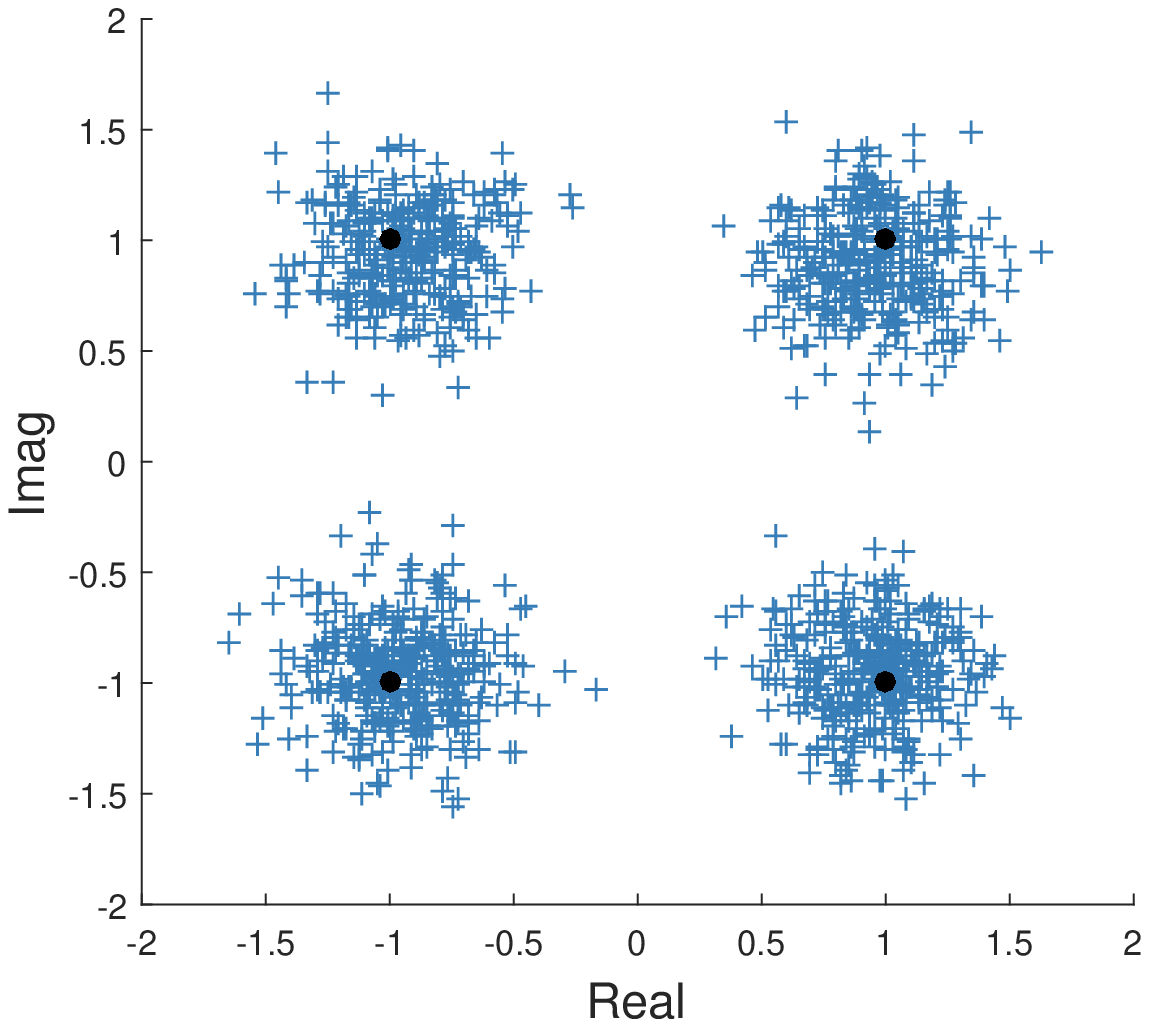}\label{fig:const-set3}
         \caption{Setup 3}
     \end{subfigure}
     \caption{Symbol constellation diagrams after equalization}
     \label{fig:const-2m_setup}
\vspace{-0.2cm}
\end{figure}

In all the experiments, 4 kbps data rates were achieved with average BER $<$2e-4, which could be brought to sufficiently low error rates for a given application with FEC. The constellation diagrams after equalization can be seen in Fig. \ref{fig:const-2m_setup}. Setup 1 provided better SNR at the output of the equalizer compared to the Setup 2 although the distance and the orientation were the same, demonstrating the importance of the multipath characteristics of the channel. In Setup 2, it was expected to have several signal arrivals with similar power due to the reflections from the table, whereas in Setup 1 the direct path was significantly different from the subsequent paths. Setup 2 provided better output SNR compared to Setup 3, because the received signal power was lower at the sides of the device compared to the front.

\subsection{Distance}
The third set of experiments were performed with the laptops placed on a table at different distances varying from 0.5 to 5 meters, and both devices transmitted and received information. The transmission was repeated twice for each distance. Device 2 (D2) had its speakers underneath, whereas Device 1 (D1) had them above the keyboard. This caused stronger direct arrival of the signal to D2 at the same sound level. Therefore, the sound levels of D1 and D2 were kept at 65\% and 80\%, respectively, to account for the different speaker configurations. 

\begin{table}[b]
    \centering
    \begin{tabular}{|l|c|c||c|c|}
    \hline
        \multirow[t]{2}{*}{\textbf{Distance}} & \multicolumn{2}{c||}{\textbf{D1} $\rightarrow$ \textbf{D2}}   & \multicolumn{2}{c|}{\textbf{D2} $\rightarrow$ \textbf{D1}} \\\cline{2-5}
                          &  {SL} & {BER}   & {SL} & {BER} \\\hline
        0.5 m      &   65\%    &  $<$2e-4      & 80\%    & $<$2e-4     \\
        2 m               &   65\%    &  $<$2e-4      & 80\%    & $<$2e-4     \\
        3 m               &   65\%    &  0.08         & 80\%    & $<$2e-4     \\
        5 m               &   65\%    &  0.1          & 80\%    & 0.09        \\
        5 m               &  100\%    &  $<$2e-4      & 100\%    & 0.01        \\\hline
    \end{tabular}
    \caption{BERs for different distances and sound levels (SLs)}
    \label{tab:2-distance-BER}
\end{table}

The resulting BERs are displayed in Table \ref{tab:2-distance-BER}. The transmission performance from D2 to D1 ({D2} $\rightarrow$ {D1}) was worse because the location of the D2's speakers obstructed a direct path and caused lower received signal levels. Nevertheless, data rates of 4 kbps were achieved for distances up to 2 meters without any detected errors in either link, and up to 5 meters with low enough BERs that could be corrected with lightweight FEC with negligible throughput impact or even with repetition codes of length 5 to 10. 

The average SNR at the output of the equalizer when communicating over 0.5 m was 28 dB. The output SNR indicates successful decoding of higher order modulation up to 1024-QAM, which would result in data rates of 20 kbps. Higher order modulation at shorter distances was not pursued in this work because a unified framework across all distances was preferred.

\vspace{-0.4cm}
\section{Conclusion}
\label{sec:conclusion}
In applications that require the transmission of a few hundreds of bits quickly and locally, e.g. user authorization or credential exchange, high rate ($>$ 1 kbps) acoustic communication provides the opportunity to initiate a connection within a fraction of a second. The NUSC systems in the literature that use built-in microphones and speakers typically operate at lower data rates because the systems and methods chosen for communication are not tailored to the properties of the channel. 

In this work, a high data rate NUSC system using built-in microphones and speakers of computers is presented. The system is designed by examining the properties of the channel and by using quadrature amplitude modulation and phase-coherent decision feedback equalization to address the challenges arising from the properties of the channel. The experiments demonstrated NUSC with data rates of 4 kbps at distances up to 5 meters with low enough error rates to be corrected with FEC. The SNR at the output of the equalizer indicated even higher data rates at shorter distances. The system proved robust for various effects such as different device orientations, coexistence with ultrasonic occupancy sensors and slow time variations due to users moving in the experiment site. Further work is necessary to analyze the effects of some other detriments (e.g., excessive movement, in-band interference) and to extend the method to other devices such as mobile phones and TVs.
\vspace{-0.3cm}
\section*{Acknowledgment}
\vspace{-0.2cm}
This work is conducted in part during an internship at Intel Corporation. The authors would like to thank Timothy Leighton for our discussions on ultrasound safety and Matias Almada for his help in field experiments.
\vspace{-0.3cm}
\section*{Appendix}
\vspace{-0.2cm}
We measured sound levels originating from the laptop at 100\% volume in the 18-20 kHz band with a measurement microphone (Behringer ECM8000) in a quiet laboratory. The noise floor was at 48.8 dB(Z) SPL, and the signal levels measured at 20 cm distance from the speaker were at 64.1, 64.3 and 62.9 dB(Z) SPL for 18, 19 and 20 kHz tones, respectively. The signal was inaudible to the experimenters. More work is needed to assess if these sound levels would comply with the guidelines or if they would cause adverse effects for some people. 

\bibliographystyle{IEEEbib-abbrv}
\bibliography{IEEEabrv,refs}

\end{document}

%% file: misc_packages.tex
\usepackage{graphicx}
\usepackage{setspace}
\usepackage{amsmath}
\usepackage{amssymb}
\usepackage{amsfonts}
\usepackage{stmaryrd}
\usepackage{color,multirow,hyperref,balance,url}
\usepackage{caption}
\usepackage{subcaption}
\usepackage{tikz}
\usetikzlibrary{shapes}
\usepackage{acro}
\acsetup{first-style=short}
\usepackage{array}
\newcolumntype{x}[1]{>{\centering\arraybackslash\hspace{0pt}}p{#1}}

\usepackage{color}

\newcounter{Q}
\newcounter{TD}

\graphicspath{{Figures/}}

\DeclareGraphicsExtensions{.pdf,.jpeg,.png}

\usepackage{epstopdf}

\usepackage{amsmath}